\documentclass[12pt]{iopart}
\usepackage{iopams}  
\usepackage{graphicx}
\def\be{\begin{equation}}
\def\ee{\end{equation}}

\newcommand{\beq}{\begin{equation}}
\newcommand{\eeq}{\end{equation}}
\newcommand{\ber}{\begin{eqnarray}}
\newcommand{\eer}{\end{eqnarray}}
\newcommand{\berr}{\begin{eqnarray*}}
\newcommand{\eerr}{\end{eqnarray*}}

\newcommand{\sg}{\sigma(\omega,\vec{p})}
\newcommand{\sgh}{\sigma_H(p_0,\vec{p})}

\begin{document}

\title{ 
Quarkonia correlators and spectral functions
from lattice QCD.
}

\author{P\'eter Petreczky 
}
\address{ 
Department of Physics and RIKEN-BNL Research Center, Brookhaven
National Laboratory, Upton, New York, 11973
}

\begin{abstract}
I discuss recent progress in calculating
quarkonia correlators and spectral functions on the
lattice in relation with the problem of quarkonia dissolution
at high temperatures and heavy quark transport in Quark Gluon Plasma.
\end{abstract}




%
\section{Introduction}
\label{intro}

Heavy quarkonia  play an important role in studying hot and dense 
strongly interacting matter. Because of the heavy quark mass quarkonia binding can be understood
in terms of the static potential. 
General considerations suggest  that quarkonia could melt at
temperatures above the deconfinement temperature
as a result of modification of inter-quark forces (color screening).
It has been conjectured by that melting of
different quarkonia states due to color screening 
can signal Quark Gluon Plasma formation in heavy ion collisions \cite{MS86}.
Many studies  of quarkonia dissolution 
rely heavily on potential models 
\cite{karsch88,digal01a,digal01b,shuryak04,wong04,alberico}.
However it is very unclear if such models are valid 
at finite temperature  \cite{petreczkyhard04}. 

The problem of quarkonium dissolution can be studied more 
rigorously in terms of meson (quarkonium) spectral functions.
Lattice calculation of charmonium spectral functions
appeared recently and suggested, contrary to potential models,
that $J/\psi$ and $\eta_c$ survive at temperatures as high as $1.6T_c$ 
\cite{umeda02,asakawa04,dattalat02,datta04}. It has been also found that $\chi_c$ 
melts at temperature of about  $1.1T_c$ \cite{dattalat02,datta04,jhw05}. 
There are also preliminary calculations of the bottomonium
spectral functions \cite{lat05,panic05}.

\section{Meson correlators and spectral functions}%

In  lattice QCD we calculate correlators of point
meson operators of the form 
\begin{equation}
	J_H(t,x)=\bar q(t,x) \Gamma_H q(t,x),
\end{equation}
where $\Gamma_H=1,\gamma_5, \gamma_{\mu}, \gamma_5 \gamma_{\mu}, 
\gamma_{\mu} \gamma_{\nu}$ 
and fixes the quantum number of the channel to 
scalar, pseudo-scalar, vector, axial-vector and tensor channels correspondingly.
The relation of these quantum number channels to different meson states is given
in Tab. \ref{tab.channels}.

\begin{table}
\begin{tabular}
[c]{||c|c|c||c|}\hline
$\Gamma$ & $^{2S+1}L_{J}$ & $J^{PC}$ & $u\overline{u}$\\\hline
$\gamma_{5}$ & $^{1}S_{0}$ & $0^{-+}$ & $\pi$\\
$\gamma_{s}$ & $^{3}S_{1}$ & $1^{--}$ & $\rho$\\
$\gamma_{s}\gamma_{s^{\prime}}$ & $^{1}P_{1}$ & $1^{+-}$ & $b_{1}$\\
$1$ & $^{3}P_{0}$ & $0^{++}$ & $a_{0}$\\
$\gamma_{5}\gamma_{s}$ & $^{3}P_{1}$ & $1^{++}$ & $a_{1}$\\
\hline
\end{tabular}%
\begin{tabular}
[c]{|cc|}\hline
$c\overline{c}(n=1)$ & $c\overline{c}(n=2)$\\\hline
$\eta_{c}$ & $\eta_{c}^{^{\prime}}$\\
$J/\psi$ & $\psi^{\prime}$\\
$h_{c}$ & \\
$\chi_{c0}$ & \\
$\chi_{c1}$ & \\
\hline
\end{tabular}
\begin{tabular}[c]{|cc|}\hline
$b\overline{b}(n=1)$ & $b\overline{b}(n=2)$\\
\hline
$\eta_b$ & $\eta_b'$ \\
$\Upsilon(1S)$ & $\Upsilon(2S)$\\
$h_b$ & \\
$\chi_{b0}(1P)$& $\chi_{b0}(2P)$\\
$\chi_{b1}(1P)$& $\chi_{b1}(2P)$\\
\hline
\end{tabular}
\caption{Meson states in different channels
\label{tab.channels}}
\end{table}

Most dynamic properties of a finite temperature system are incorporated 
in the spectral function. The spectral function $\sgh$ for a given 
mesonic channel $H$ in a system at temperature $T$ can be defined 
through the Fourier transform of the real time two-point functions
$D^{>}$ and $D^{<}$ or, equivalently, as the imaginary part of 
the Fourier transformed retarded 
correlation function \cite{lebellac},
\ber
\sgh &=& \frac{1}{2 \pi} (D^{>}_H(p_0, \vec{p})-D^{<}_H(p_0, \vec{p}))=
\label{eq.defspect}\\
&&
\frac{1}{\pi} Im D^R_H(p_0, \vec{p}) \nonumber \\
 D^{>(<)}_H(p_0, \vec{p}) &=& \int{d^4 p \over (2
\pi)^4} e^{i p \cdot x} D^{>(<)}_H(x_0,\vec{x}) \\
D^{>}_H(x_0,\vec{x}) &=& \langle
J_H(x_0, \vec{x})~ J_H(0, \vec{0}) \rangle \nonumber\\
D^{<}_H(x_0,\vec{x}) &=& 
\langle J_H(0, \vec{0})~ J_H(x_0,\vec{x}) \rangle , \quad x_0>0 \
\eer

The Euclidean time correlator calculated on the lattice
\beq
G_H(\tau, \vec{p}) = \int d^3x e^{i \vec{p} \cdot \vec{x}} 
\langle T_{\tau} J_H(\tau, \vec{x}) J_H(0,
\vec{0}) \rangle
\eeq
is an analytic continuation of the real time correlator
$G_H(\tau,\vec{p})=D^{>}(-i\tau,\vec{p})$. 

Using this equation and the  Kubo-Martin-Schwinger
(KMS) condition \cite{lebellac} for the correlators
\beq
D^{>}_H(x_0,\vec{x})=D^{<}(x_0+i/T,\vec{x}),
\label{kms}
\eeq
one can relate the Euclidean propagator $G_H(\tau,\vec{p})$ to the 
spectral function in Eq. (\ref{eq.defspect}), through the integral
representation
\ber
G(\tau, \vec{p}) &=& \int_0^{\infty} d \omega
\sg K(\omega, \tau) \label{eq.spect} \nonumber\\
K(\omega, \tau) &=& \frac{\cosh(\omega(\tau-1/2
T))}{\sinh(\omega/2 T)}.
\label{eq.kernel}
\eer

To reconstruct the spectral function from the lattice correlator 
$G(\tau,T)$ this integral representation should be inverted. 
Since the number of data points is less than the number of degrees
of freedom (which is ${\cal O}(100)$ for reasonable discretization of
the integral ) spectral functions can be reconstructed only using the
Maximum Entropy Method (MEM) \cite{asakawa01}. 
In this method one looks for a spectral function which
maximizes the
conditional probability $P[\sigma|DH]$ of having the spectral function $\sigma$ given
the data $D$ and some prior knowledge $H$ which for positive definite spectral function
can be written as 
\beq
 P[\sigma|DH]=\exp(-\frac{1}{2} \chi^2 + \alpha S),
\label{eq:PDH}
\eeq
where 
\beq
S=\int d \omega \biggl [ \sigma(\omega)-m(\omega)-\sigma(\omega)
  \ln(\frac{\sigma(\omega)}{m(\omega)}) \biggr]
\eeq
is the Shannon - Janes entropy. 
The real function $m(\omega)$ is called the default model and parametrizes all
additional prior knowledge about the
spectral functions, such as the asymptotic behavior at high energy  \cite{asakawa01}.
In order to have sufficient
number of data points either very fine isotropic lattices \cite{dattalat02,datta04,panic05}
or anisotropic lattices \cite{umeda02,asakawa04,jhw05,lat05} have been used.

\section{Charmonia correlators and spectral functions}

The spectral function for pseudo-scalar charmonium spectral functions calculated on 
anisotropic lattice \cite{jhw05} is shown in Fig. 1.
The first peak in the spectral function corresponds to $\eta_c(1S)$ state.  The position of the
peak and the corresponding amplitude (i.e. the area under the peak) are in good agreement
with the results of simple exponential fit. The second peak in the spectral function is most likely
the combination of several excited states as its position and amplitude is higher than what one would
expect for pure 2S state. The spectral function becomes sensitive to the effects of the finite lattice spacings
for $\omega>5$GeV. In this $\omega$ region the spectral functions becomes also sensitive to the choice
of the default model. This is because only a very few data points in the correlator carry information 
about the spectral function in the region  $\omega>5$GeV. 

Also shown in Fig.1 is the spectral function in the
scalar channel from Ref. \cite{jhw05}. 
The 1st peak corresponds to $\chi_{c0}(1P)$ state. The correlator is more noisy
in the scalar channel than in the pseudo-scalar one. As the results the $\chi_{c0}(1P)$ peak
is less pronounced and has larger statistical errors. The peak position and the area under the peak
is consistent with the simple exponential fit. 
As in the pseudo-scalar case individual excited states are
not resolved and the spectral function depends on the lattice spacing and default model for $\omega>5$GeV. 
\begin{figure}
	\includegraphics[width=7cm]{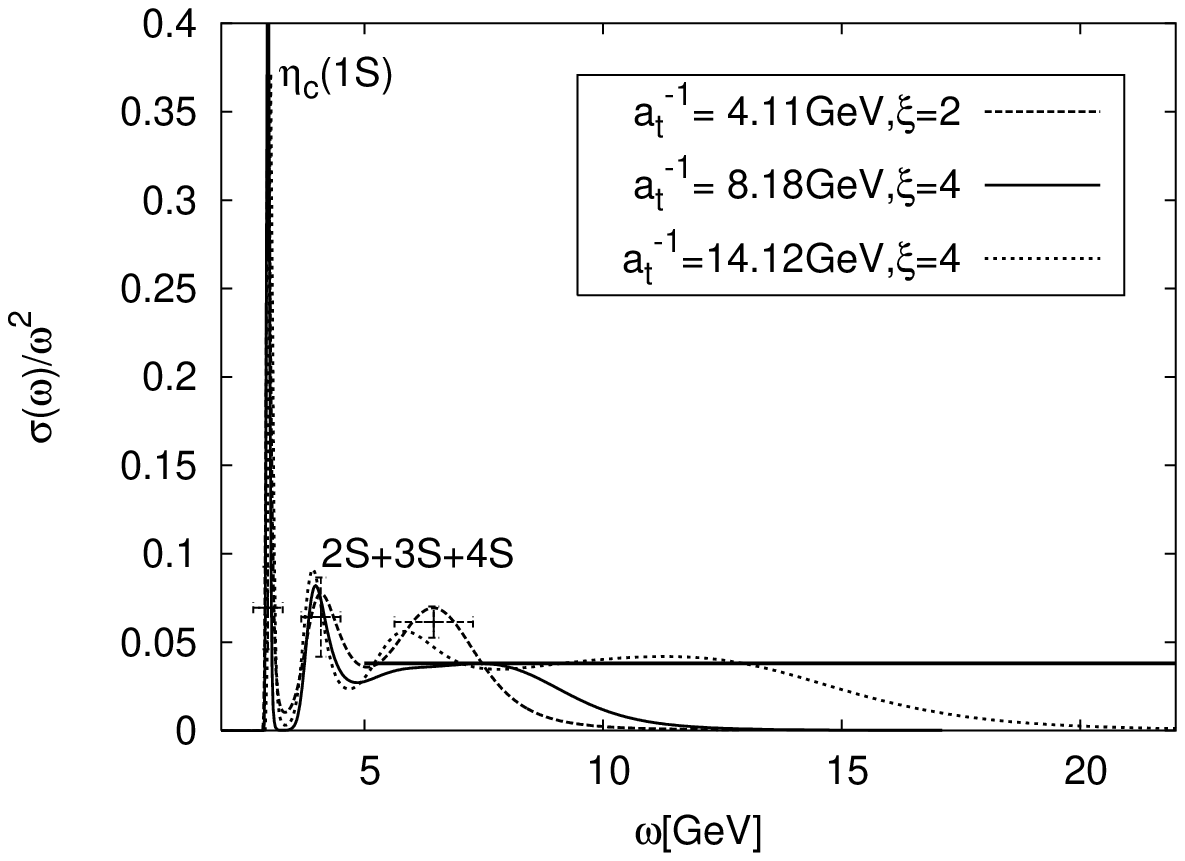}
	\includegraphics[width=7cm]{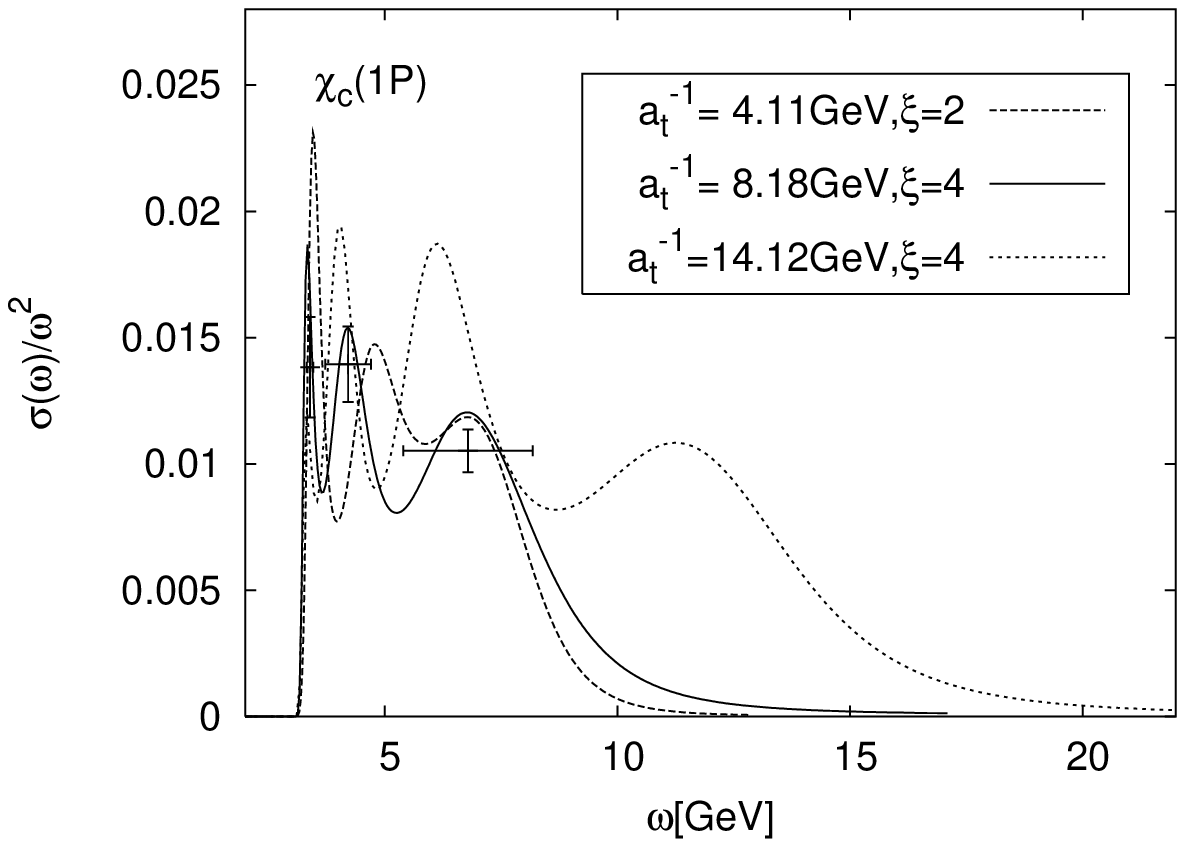}
\caption{
Charmonium spectral function in  the pseudo-scalar channel (left) and 
the scalar channel (right) at different lattice spacings and zero temperature from Ref. \cite{jhw05}.
$m(\omega)=1$ was used as the default model.}
\label{fig.spfct0}
\end{figure}
Similar results have been found for the vector and axial-vector
channels which correspond to $J/\psi$ and $\chi_{c1}$ states respectively.

\begin{figure}
\includegraphics[width=3in]{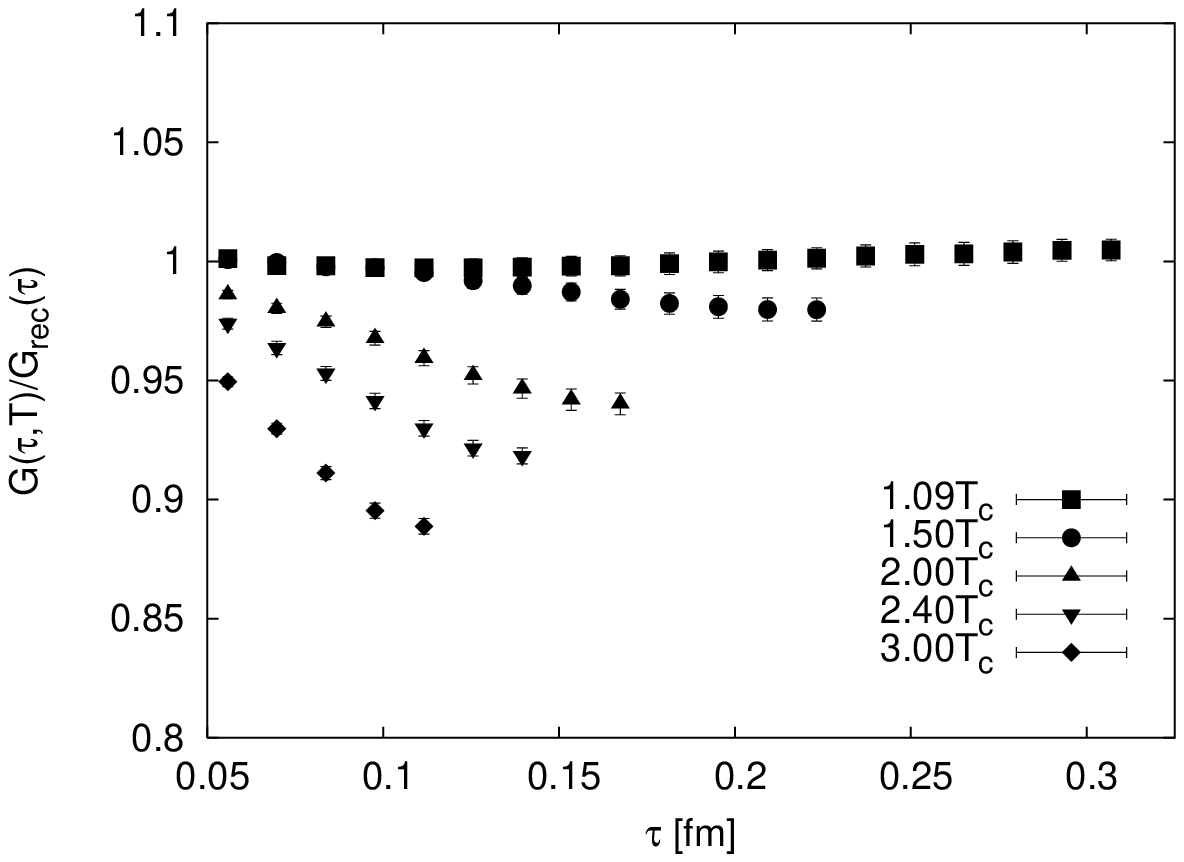}
\includegraphics[width=3in]{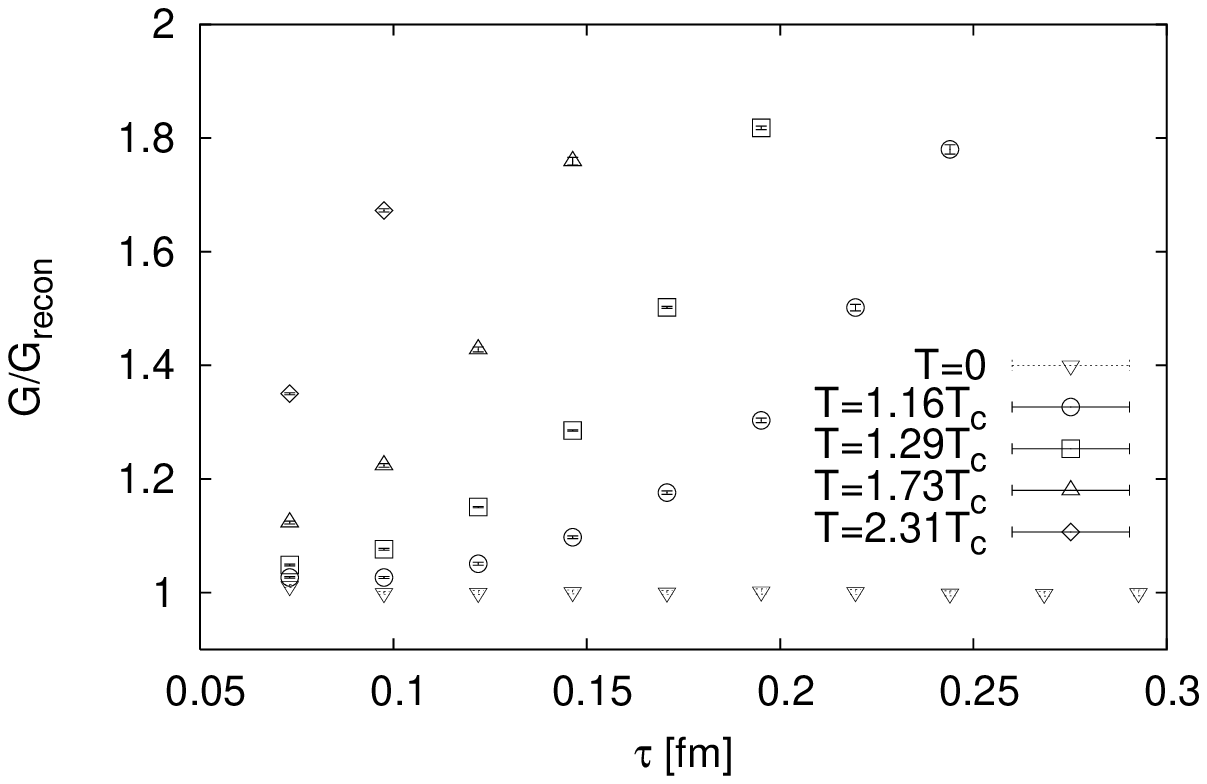}
\caption{The ratio $G(\tau,T)/G_{recon}(\tau,T)$ of charmonium for pseudoscalar channel at  at $a_t^{-2}=14.11$GeV
(left) and scalar channel at  at $a_t^{-2}=8.18$GeV  (right) at different
temperatures \cite{jhw05}.}
\label{ratioc} 
\end{figure}

We would like to know what happens to different charmonia states
at temperatures above the deconfinement temperature $T_c$. With
increasing temperature it becomes more and more difficult to
reconstruct the spectral functions as both the number of available
data points as well as the physical extent of the time direction
(which is $1/T$) decreases. Therefore it is useful to study the
temperature dependence of charmonia correlators first. From
Eq. (\ref{eq.kernel}) it is clear that the temperature dependence 
of charmonia correlators come from two sources: the temperature
dependence of the spectral function and the temperature dependence of
the integration kernel $K(\tau,\omega,T)$. To separate out the 
trivial temperature dependence due to the integration kernel,
following Ref. \cite{datta04} for each temperature we calculate
the so-called reconstructed correlator  
\begin{equation}
G_{recon}(\tau,T)=\int_{0}^{\infty}d\omega
\sigma(\omega,T=0)K(\tau,\omega,T).
\end{equation}
Now if we assume that there is no temperature dependence 
in the spectral function, then the ratio of the original and 
the reconstructed correlator should be close to one,
$G(\tau,T)/G_{recon}(\tau,T) \sim 1$. 
This way we can identify the cases when spectral function itself 
changes dramatically with temperature. 
This gives reliable information about the fate of charmonia states above
deconfinement. 
In Fig.~\ref{ratioc} we show this ratio for pseudo-scalar and scalar 
channels correspondingly calculated on anisotropic lattice \cite{jhw05}. 
From the figures one can see that the pseudo-scalar correlators shows
only very small changes till $1.5T_c$ indicating that the $\eta_c$ state survives
till this temperature with little modification of its properties. On the other hand the scalar
correlator shows large changes already at $1.16T_c$ suggesting strong modification or
dissolution of the  $\chi_{c0}$ state at this temperature.

\begin{figure}
	\includegraphics[width=7cm]{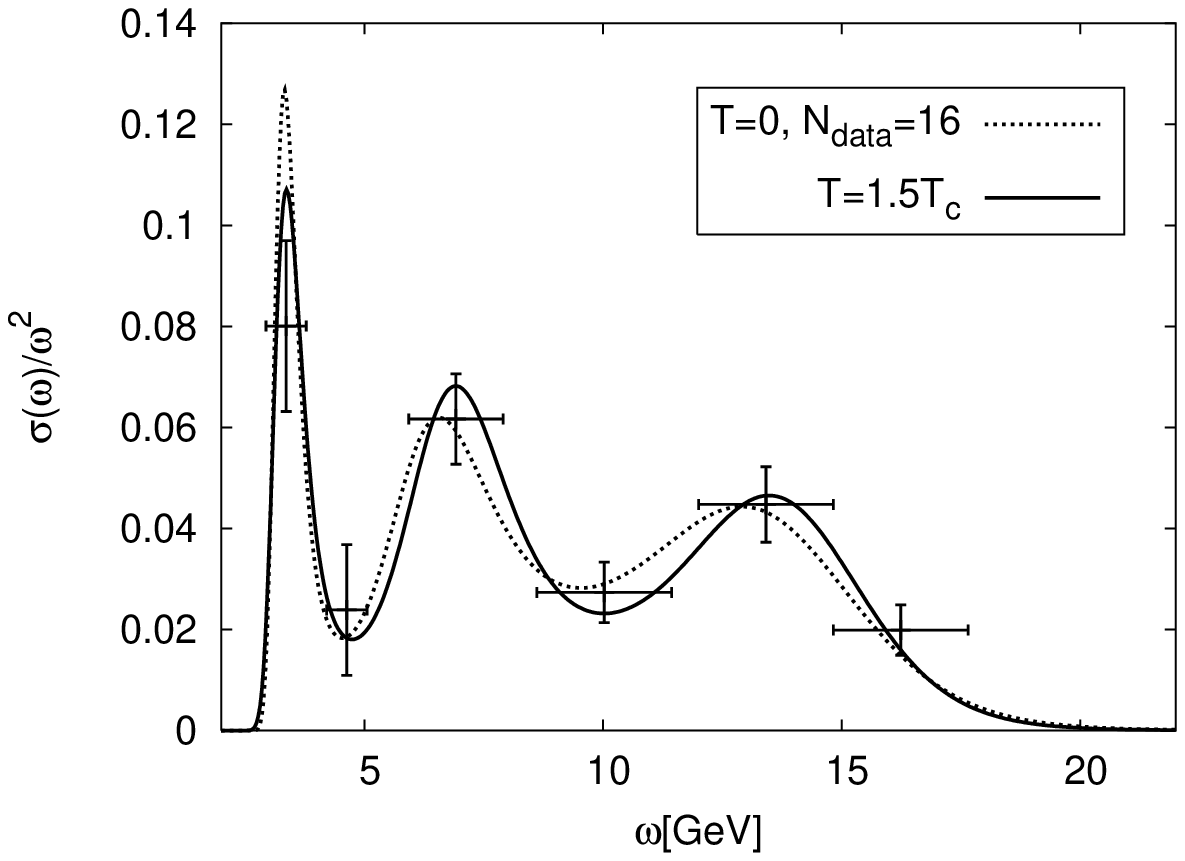}
	\includegraphics[width=7cm]{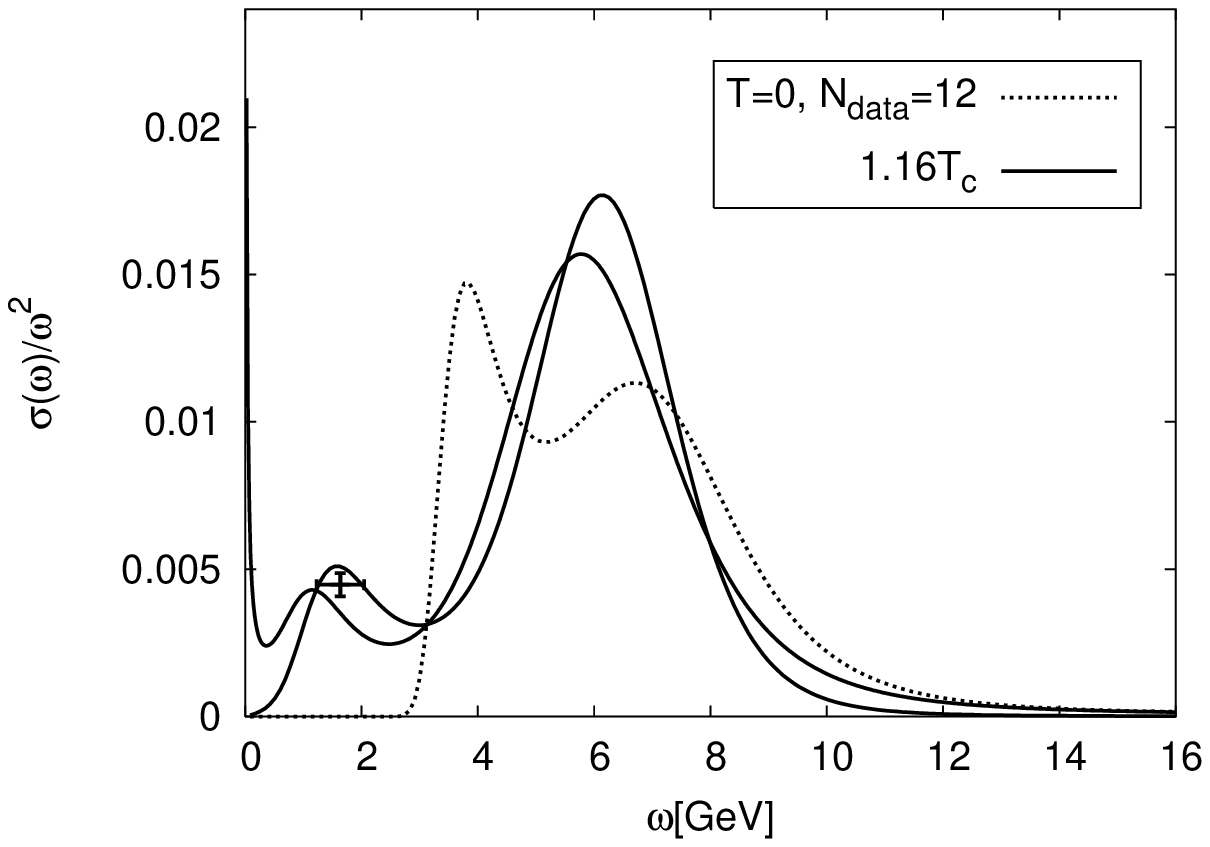}
\caption{
Charmonia spectral function in  the psuedoscalar channel at $a_t^{-2}=14.11$GeV  (left) and 
the scalar channel (right)   at $a_t^{-2}=8.18$GeV  for zero and finite temperature \cite{jhw05}. 
In the pseudo-scalar channel $m(\omega)=0.1$ has been used as the default model. In the scalar channel
we have used $m(\omega)=0.038\omega^2$ as the default model as well as $m(\omega)=1$ for $T=1.16T_c$.}
\label{fig.spfct}
\end{figure}

More detailed information on different charmonia
states at finite temperature can be obtained by
calculating spectral functions using MEM.
The results of these calculations are show in
Figs.~\ref{fig.spfct}. 
Because at high temperature the temporal extent and the number of data
points where the correlators are calculated become smaller the spectral functions reconstructed
using MEM are less reliable.  To take into account possible systematic 
effects when studying  the temperature modifications of the spectral functions
we compare the finite temperature spectral functions against the zero temperature spectral functions
obtained from the correlator  using the same time interval and number of data points as available
at finite temperature. 
We see that spectral function in the pseudo-scalar channel show no temperature dependence
within the statistical errors.
This is in accord with the analysis of the correlation functions.
Also the spectral functions show very little dependence on the default model. 
Similar conclusion has been made in Ref. \cite{dattalat02,datta04} where correlators and
spectral functions have been calculated on very fine isotropic lattices as well as in
Ref. \cite{asakawa04} where anisotropic lattice have been used. The pseudo-scalar spectral function
was found to be temperature independent also in Ref. \cite{umeda02} where 
correlators of 
extended meson operators have been studied on anisotropic lattices. The study of the charmonium correlators
with different spatial boundary conditions provides further evidence for
survival of the $1S$ charmonia states well above the deconfinement transition
temperature \cite{iida}.

The scalar spectral function shows large changes at
$1.16T_c$ which is consistent with correlator-based analysis. Also default model dependence
of the scalar correlator is large above the deconfinement transition (c.f. Fig. 3, right).
This means that the $\chi_{c0}$ ($^3P_0$) dissolves at this temperature. 
Similar results for the scalar spectral function have been reported in \cite{dattalat02,datta04}.
The results for the axial correlators and spectral functions are similar to scalar ones 
\cite{datta04} as expected. 

The vector correlator, however, has temperature dependence different
from that of the pseudo-scalar channel \cite{mylat05}. 
This is due to the fact that the vector current is
conserved and there is a contribution to spectral functions at very small 
energy $\omega \simeq 0$
corresponding to heavy quark transport \cite{derek,mocsyhard04}.
The transport peak in the spectral functions can be written as \cite{derek}
\begin{equation}
\sigma_{low}(\omega)=\chi_s(T) \frac{T}{M} \frac{1}{\pi}\frac{\omega \eta}{\omega^2+\eta^2},
\end{equation}
where $\eta=T/(MD)$ with $D$ being the heavy
quark diffusion constant. Furthermore $\chi_s(T)$ is the charm or beauty 
susceptibility and $M$ is the heavy quark mass.
To the first approximation the transport contribution to the spectral
function gives rise to a positive constant contribution to the correlator 
$G_{low}(\tau) \simeq \chi_s(T) T^2/M$ \cite{derek} resulting in an enhancement 
of the finite temperature correlators relative to the zero temperature ones, in agreement with the lattice 
data presented in \cite{mylat05,dattasewm04}. Finite value of the diffusion
constant $D$ will give rise to some curvature in $G_{low}(\tau)$. The smaller
the value of $D$ is,  the larger is the curvature in $G_{low}(\tau)$. Thus extracting
$G_{low}(\tau)$ from lattice data and estimating its curvature can give an estimate
for $D$.
This, however, requires very precise lattice data which are not yet available \cite{derek}.

\section{Bottomonium correlators and spectral functions}

\begin{figure}
\includegraphics[width=3in]{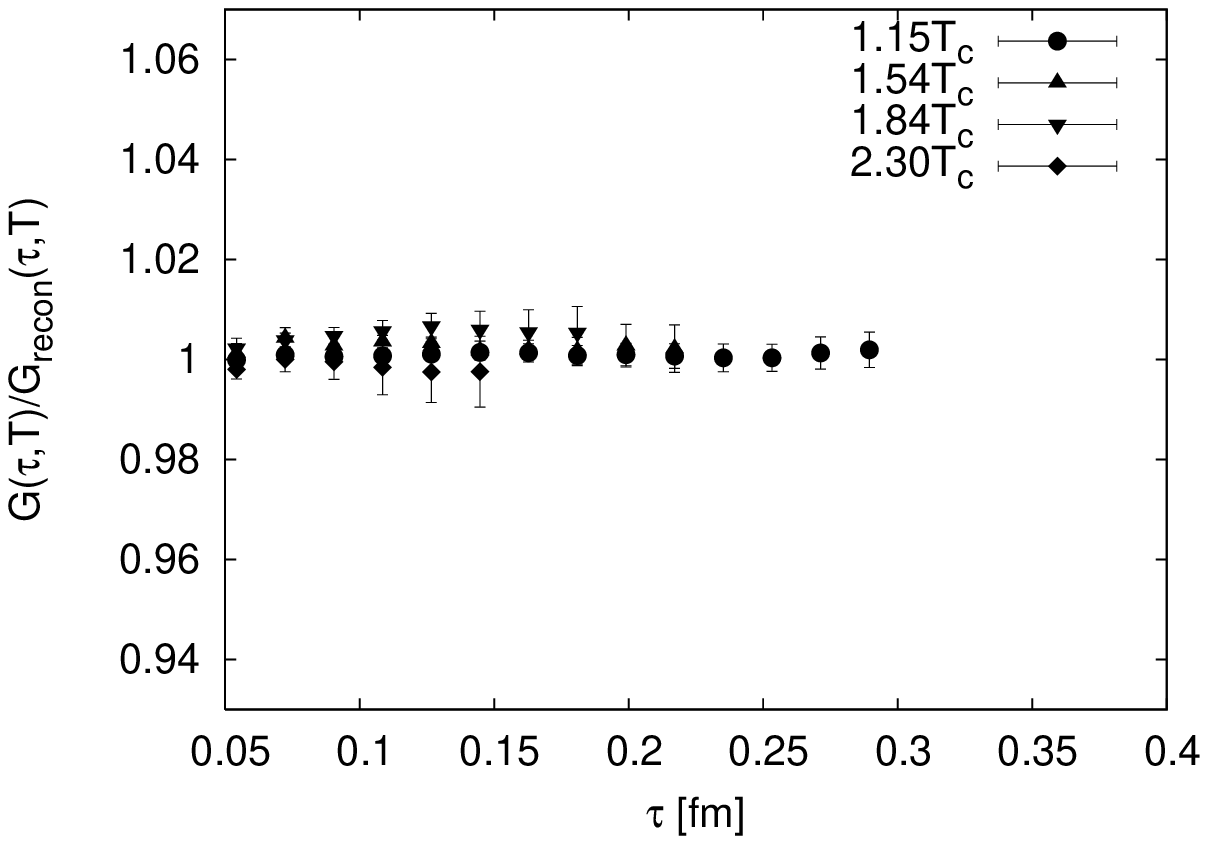}
\includegraphics[width=3in]{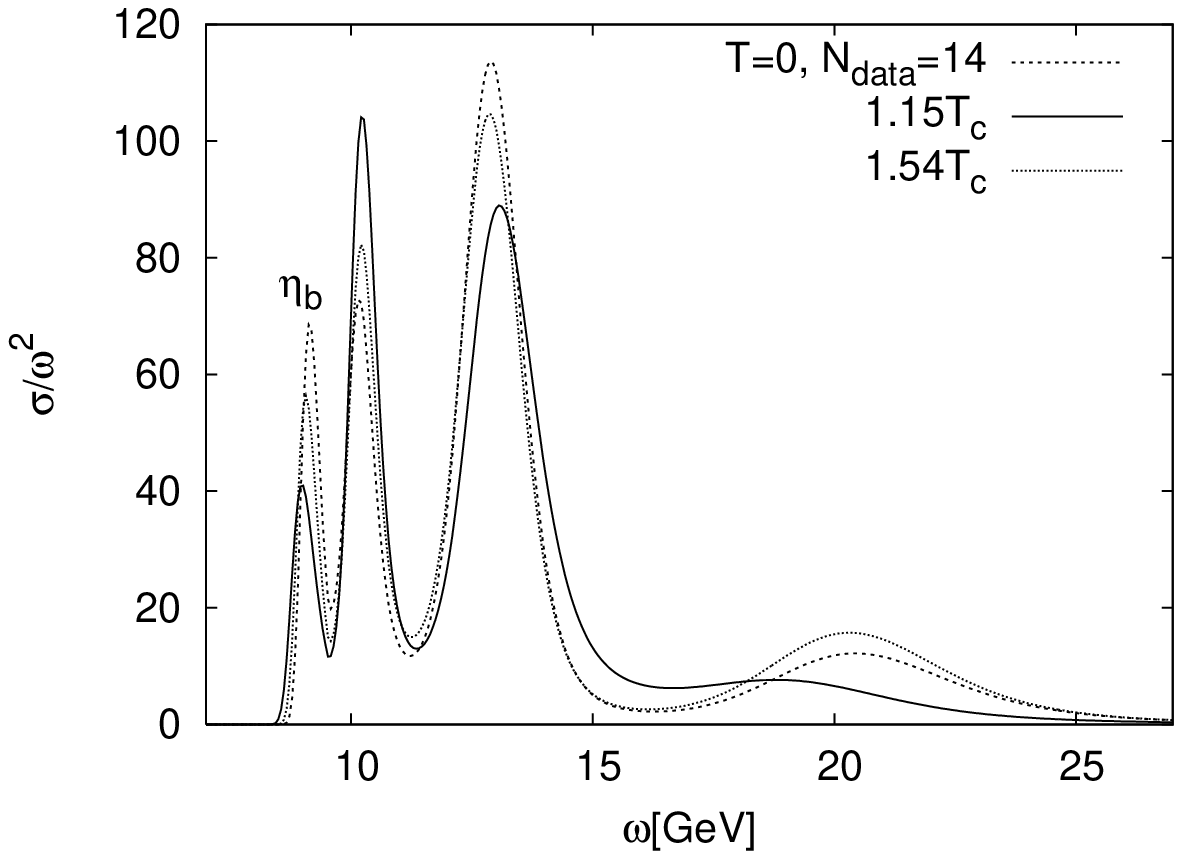}
\caption{Bottomonia correlators (left) spectral functions (right) 
in pseudo-scalar
channel for different temperatures from Ref. \cite{lat05}.
}
\label{psbot}
\end{figure}

Bottomonium correlators and spectral functions have also been studied using 
anisotropic \cite{lat05} as well as very fine isotropic lattice with lattice spacing
$a^{-1}=9.72$GeV \cite{panic05}.
These studies, however,
are far less detailed than the charmonium studies presented above and are still preliminary.
In Fig. \ref{psbot} the temperature dependence of the ratio $G/G_{recon}$ is shown for pseudo-scalar
channel at different temperatures calculated using anisotropic lattices. 
This ratio shows almost no temperature dependence 
till $2.3T_c$. This
is expected as the pseudo-scalar $1S$ bottomonium state, the $\eta_b$,  is much 
smaller than the corresponding charmonium state, thus survives till much higher temperatures.
Also shown in Fig. \ref{psbot} is the pseudo-scalar function at different temperatures. 
The first peak corresponds to the $\eta_b$ state and survives above the
deconfinement temperature in agreement with the analysis of the correlation functions. Other 
peaks are artifacts of the finite
lattice spacing and MEM analysis.

\begin{figure}
\includegraphics[width=3in]{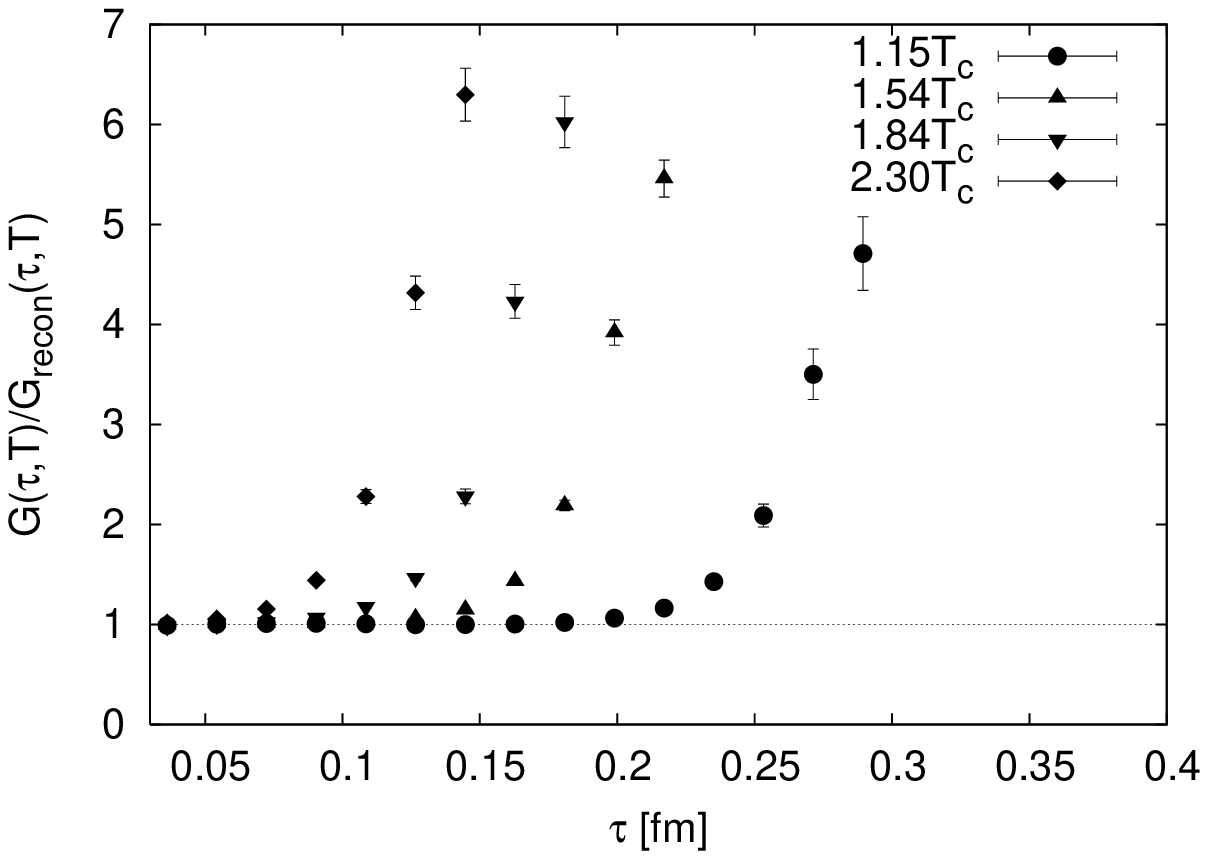}
\includegraphics[width=3in]{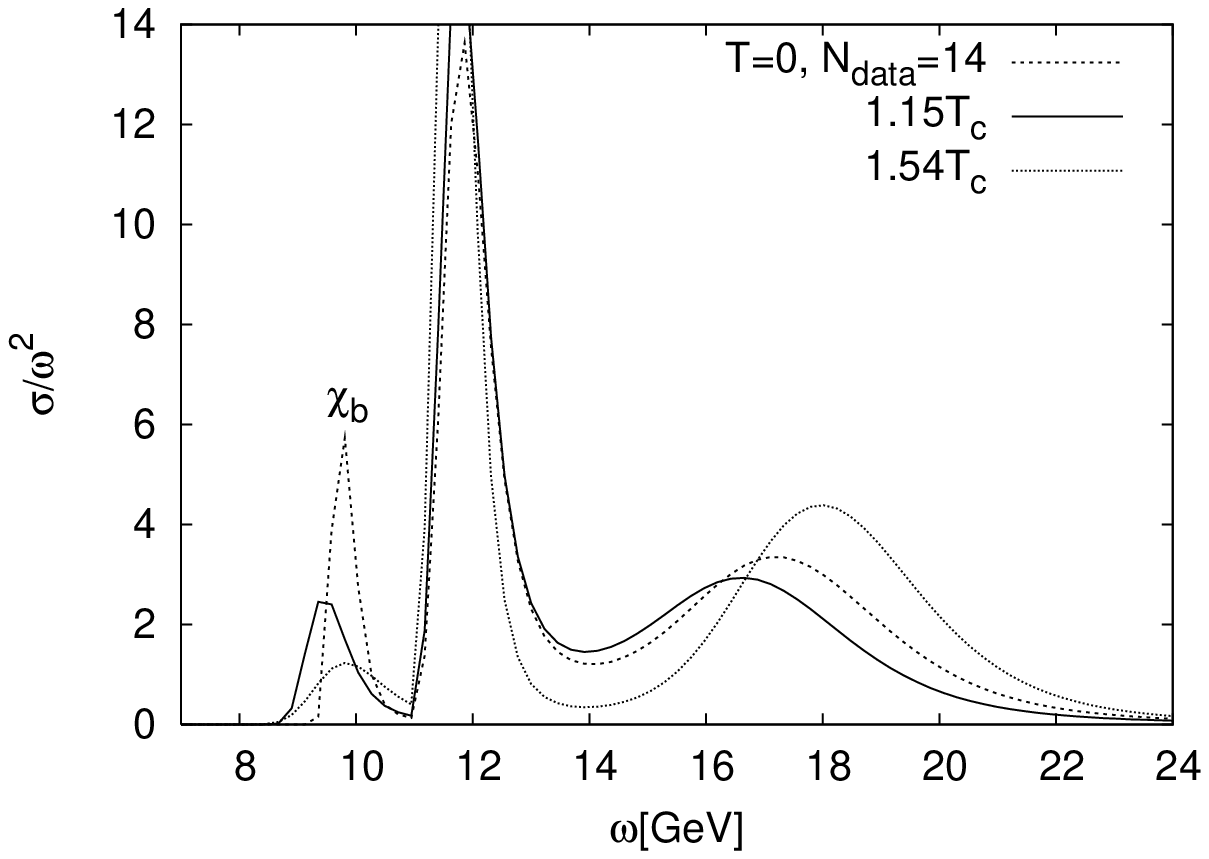}
\caption{Bottomonia correlators (left) spectral functions (right) 
in scalar
channel for different temperatures from Ref. \cite{lat05}.
}
\label{scbot}
\end{figure}

An interesting question is what happens to $1P$ bottomonia states. They have sizes 
similar to the $1S$ charmonia states and thus are expected to survive in the deconfined
phase till temperatures of about $1.5T_c$. In Fig. \ref{scbot} I show the temperature 
dependence of $G/G_{recon}$ in the scalar channel corresponding to $\chi_{b0}$. 
As one can see from the figure the scalar correlator shows dramatic change across the
deconfinement temperature and its behavior is similar to the behavior of the scalar correlator 
in the charmonium case. Also shown in Fig. \ref{scbot} is the bottomonium spectral functions
in the scalar channel. Contrary to the pseudo-scalar channel the scalar spectral function shows
significant changes above the deconfinement temperature and therefore it seems that the $\chi_b$ state is strongly modified or
even dissolved at $1.15T_c<T<1.5T_c$.
Lattice calculations presented in Ref. \cite{panic05} show similar results. In particular,
they also show large increase of the scalar bottomonium correlator and strong modification
of the corresponding spectral function.

\section{Conclusions}

In this contribution it has been shown how lattice calculation can provide
information on quarkonia properties at finite temperature. The temperature 
dependence of the pseudo-scalar correlators as well as the spectral functions
extracted using MEM shows that the $1S$ charmonia states exist as a resonance 
in the deconfined phase till temperatures as high as $1.5T_c$. On the other
hand lattice calculation show that the $1P$ charmonia states 
dissolve at $T \sim 1.1T_c$. Study of the bottomonia at finite temperature is
also in progress \cite{lat05,panic05}. It should be stressed that all lattice 
calculation discussed so far have been done in quenched approximation, i.e.
neglecting the effect of sea quarks. To make contact with heavy ion experiments
certainly the effect of the sea quarks has to be included, but computationally this is very 
expensive. Recent attempts to study charmonia properties at
finite temperature in full QCD (i.e. with sea quarks) are reported in Ref. \cite{skull}.
The findings of Ref. \cite{skull} are consistent with the quenched results.

Recent studies of quarkonium properties at finite temperature using potential 
model claim agreement with the lattice data \cite{shuryak04,wong04,alberico}. 
The dissociation temperature of different
quarkonia states quoted in these studies and 
defined as temperature where binding energy becomes zero
is indeed significantly higher than before. 
However, potential models with screened potential also 
predict modification of quarkonia properties, which in turn
lead to changes in the spectral function and correlators.
Such changes 
are not observed in the lattice correlator \cite{mocsyhard04,mocsythis}.
Thus it is not clear at the moment whether modification of quarkonia
properties at finite temperature can be understood in terms
of a screened temperature dependent heavy quark potential.

\section{Acknowledgments} 
This work was supported by U.S. Department of Energy under 
Contract No. DE-AC02-98CH10886 as well as by LDRD funds.

\end{document}